\documentstyle[preprint,prb,aps]{revtex}
\def\rr{{{\bf r}}}
\def\rrp{{{\bf r^\prime}}}

\begin{document}
\draft

\title{ Quantum size correction to the work function and
the centroid of excess charge in positively ionized simple metal clusters}

\author{M. Payami }
\address{Center for Theoretical Physics and Mathematics,
Atomic Energy Organization of Iran,\\ P.~O.~Box 11365-8486, Tehran, Iran}

\date{\today}
\maketitle
\begin{abstract}

In this work, we have shown the important role of the finite-size correction
to the work function in predicting the correct positions
of the centroid
of excess charge in positively charged simple metal clusters with different
$r_s$ values ($2\le r_s\le7$). For this purpose, firstly we have calculated
the self-consistent Kohn-Sham energies of neutral and singly-ionized clusters
with sizes $2\le N\le 100$ in the framework of local spin-density approximation
and stabilized jellium model (SJM) as well as simple jellium model (JM) with
rigid
jellium. Secondly, we have fitted our results to the asymptotic ionization
formulas both with and without the size correction to the work function. The
results of fittings show that the formula containing the size correction
predict
a correct position of the centroid inside the jellium while the other predicts
a false position outside the jellium sphere.

\end{abstract}
\newpage

\section{Introduction}
\label{sec1}

In the SJM,\cite{pertran} the total energy of a cluster
with jellium radius $R=\bar r_s^{1/3}N^*$ is given by

\begin{eqnarray}\label{eq1}
E_R[n]=T_s[n]+E_{xc}[n]&+&\int d\rr\;\{\frac{1}{2}\phi_R[n;\rr]
+\langle\delta v\rangle_{\rm WS}\Theta(\rr)\}[n(\rr)-n_+(\rr)] \\  \nonumber
&+& (\varepsilon_{\rm M}+\bar w_R) \int d\rr\;n_+(\rr).
\end{eqnarray}

The first and second terms in
the right hand side of Eq.(\ref{eq1}) are
the non-interacting kinetic and the exchange-correlation energies, respectively
and

\begin{equation}\label{eq2}
\phi_R([n];\rr)=\int
d\rrp\;\frac{n(\rrp)-n_+(\rrp)}{\left|\rr-\rrp\right|}.
\end{equation}

All equations throughout this paper are in atomic units ( $\hbar=e^2=m=a_0=1$)
unless otherwise explicitly expressed. $n_+(\rr)$ is the positive jellium
background density, $n_+(\rr)=\bar n \theta(R-r)$ where $\bar n=3/4\pi\bar
r_s^3$ is the bulk density.
The quantity $\langle\delta v\rangle_{\rm WS}$
is\cite{pertran} the average of the difference potential over the
Wigner-Seitz cell and the difference potential, $\delta v$, is defined as the
difference between the pseudopotential of a lattice of ions and the
electrostatic potential of the jellium positive background.
$\Theta(\rr)$
takes the value of unity inside the jellium background and zero, outside.
$\varepsilon_{\rm M}$ and $\bar w_R$ are the Madelung energy and the repulsive
part of the pseudo-potential, respectively.

In the continuum approximation,
the ground-state density, $n_R(\rr)$, of the neutral cluster is given by the
solution of the Euler equation\cite{parr}

\begin{equation}\label{eq3}
\mu(R)=\left.\frac{\delta E_R[n]}{\delta n(\rr)}\right|_{n=n_R}=
\phi_R([n_R];\rr) + \langle\delta v\rangle_{\rm WS}\Theta(\rr)+
\left.\frac{\delta E_{kxc}[n]}{\delta n(\rr)}\right|_{n=n_R}.
\end{equation}
In Eq (\ref{eq3}), $\mu(R)$ is the chemical potential of electrons which is
constant throughout the cluster for the exact ground-state density $n_R$, and
$E_{kxc}=T_s+E_{xc}$. The ground-state
density satisfies the costraint $\int d\rr\;n_R(\rr)=N^*$. Removing an
electron from the neutral cluster gives rise to a new ground-state
density $n_R^\prime(\rr)=n_R(\rr)+\delta n(\rr)$ with $\int d\rr\;\delta
n(\rr)=-1$.

The ionization energy is defined as the difference in the ground-state energies
of the two systems with $N^*$ and $(N^*-1)$ electrons which may be represented
by the expansion

\begin{eqnarray}\label{eq4}
I(R)&=&E_R[n_R^\prime]-E_R[n_R] \\  \nonumber
   &=&\int d\rr\;\left.\frac{\delta E_R[n]}{\delta n(\rr)}\right|_{n=n_R}\delta
n(\rr)  + \frac{1}{2}\int d\rr\int d\rrp\left.\frac{\delta^2 E_R[n]}{\delta
n(\rr)\delta n(\rrp)}\right|_{n=n_R}\delta n(\rr)\delta n(\rrp) + \cdots
\end{eqnarray}

The first term in the right hand side of Eq. (\ref{eq4}) is $-\mu(R)$ and
in local density approximation for the kinetic and exchange-correlation
energies we have

\begin{equation}\label{eq5}
\left.\frac{\delta^2 E_R[n]}{\delta n(\rr)\delta n(\rrp)}\right|_{n=n_R}=
\frac{1}{|\rr-\rrp|} + \delta(\rr-\rrp)\left.\frac{\partial^2}{\partial
n^2}\{n\varepsilon_{kxc}(n)\}\right|_{n=n_R(\rr)}.
\end{equation}

Here, $\varepsilon_{kxc}(n)$ is the sum of the bulk kinetic and
exchange-correlation energies per electron at the
density $n$. Choosing a spherical distribution\cite{per88} for $\delta n(\rr)$

\begin{equation}\label{eq6}
\delta n(\rr)=\frac{\xi(r)}{4\pi(R+a)^2},
\end{equation}
and interpretting $(R+a)$ as the centroid of the excess charge, $\xi(r)$
satisfies the following two equations:

\begin{equation}\label{eq7}
\int_0^\infty \frac{\xi(r)}{4\pi(R+a)^2}4\pi r^2\;dr=-1
\end{equation}

\begin{equation}\label{eq8}
\int_0^\infty r\frac{\xi(r)}{4\pi(R+a)^2}4\pi r^2\;dr=-(R+a).
\end{equation}

The simplest choice for $\xi(r)$ which stisfies the above two constraints, is

\begin{equation}\label{eq9}
\xi(r)=-\delta(r-R-a).
\end{equation}

Using the Eqs. (\ref{eq5})-(\ref{eq9}) in the second term of the right hand
side of the Eq. (\ref{eq4}) one obtains

\begin{equation}\label{eq10}
I(R)=-\mu(R)+\frac{1}{2(R+a)}+O(R^{-2}).
\end{equation}

Since $\mu(R)$ is a constant, independent of $\rr$, for the exact ground-state
density of the neutral cluster, $n_R(\rr)$, it is possible to evaluate it at a
point deep inside the cluster where $n_R(\rr)=\bar n$. The density profile
$n_R(r)$ and the potential $\phi_R([n_R],r)$ for large clusters can be
expressed as corresponding
quantities for the planar surface (i.e. $R=\infty$) plus a size
correction\cite{engper91}

\begin{equation}\label{eq11}
n_R(x)=n(x)+\frac{f(x)}{R}+O(R^{-2})
\end{equation}

\begin{equation}\label{eq12}
\phi_R(x)=\phi(x)+\frac{h(x)}{R}+O(R^{-2})
\end{equation}
where $x=r-R$ gives the distance from the jellium edge. At large distances
from the jellium edge outside the cluster, we have
$f(+\infty)=\phi(+\infty)=h(+\infty)=0$, and
for a point deep inside the cluster ($x=-\infty$) we have
$n(-\infty)=\bar n$, $f(-\infty)=0$. Hence Eq. (\ref{eq3}) reads

\begin{equation}\label{eq13}
\mu(R)=\phi(-\infty)+\frac{h(-\infty)}{R}+\langle\delta v\rangle_{\rm WS}
+\left.\frac{\partial}{\partial n}\{n\varepsilon_{kxc}(n)\}\right|_{n=\bar n}.
\end{equation}

Upon inserting Eq. (\ref{eq13}) into Eq. (\ref{eq10}) and using the definition
of the work function of a planar surface in the SJM
\cite{seidl98}

\begin{equation}\label{eq14}
W=\Delta\phi-\langle\delta v\rangle_{\rm WS}-
\left.\frac{\partial}{\partial n}\{n\varepsilon_{kxc}(n)\}\right|_{n=\bar n},
\end{equation}
where $\Delta\phi=\phi(+\infty)-\phi(-\infty)$ is the electrostatic dipole
barrier\cite{lang70}, one obtains

\begin{equation}\label{eq15}
I(R)=\left(W+\frac{c}{R}\right)+\frac{1}{2(R+a)}+O(R^{-2}),
\end{equation}
where $c=-h(-\infty)$.

Equation (\ref{eq15}) is also valid in the JM for which $W$
is obtained by putting $\langle\delta v\rangle_{\rm WS}=0$ in Eq. (\ref{eq14})
and recalculating $\Delta\phi$ with the JM density profile.
There are two methods\cite{monnier,seidl96,seidl98} for calculating $W$ and $c$
called the "Koopmans' method"
and the "change in self-consistent field ($\Delta SCF$) method" which give the
same results for the correct $n(x)$, but
 the $\Delta SCF$ method has the advantage that its results are less sensitive
to the exact density profile $n(x)$. However, in our calculations, all the
quantities $W$, $c$, and $a$ are obtained from fitting of our self-consistent
ionization energies of the clusters to Eq. (\ref{eq15}).

\section{Calculational Scheme and Results}
In this work, using the JM as well as the SJM with local spin-density
approximation,\cite{pay01} we have solved the self-consistent Kohn-Sham
equations\cite{kohnsham} and obtained the energies of
neutral and singly-ionized $N$-atomic clusters of different sizes ($2\le N\le
100$). The calculations have been performed for different $\bar r_s$ values of
the jellium sphere ($2\le\bar r_s\le 7$) in steps 0.5 bohr. For each
$\bar r_s$ value, we have calculated the ionization energies

\begin{equation}\label{eq16}
I(R)=E_R(N-1)-E_R(N),
\end{equation}
for $2\le N\le 100$ with $R=N^{1/3}\bar r_s$. Then, by fitting the $I(R)$
values
to Eq. (\ref{eq10}) we have obtained the values of $W$, $c$, and $a$ for
that $\bar r_s$ value.

Figure \ref{fig1}(a) shows the work function obtained for our SJM data and is
labeled as SJM(1) because, in this paper we have called the
Eq. (\ref{eq15}) as scheme 1. To compare
our results with the bulk calculation results , SJM(bulk), we have also plotted
the
data from table IV of Ref. \ref{fiol92}. As is seen, our results show a good
agreement with the bulk calculations results.

In Fig. \ref{fig1}(b) we have compared our JM(1) results with the bulk JM
results of Ref. \ref{fiol92}. Here also the good agreement is obvious.

Figure \ref{fig1}(c) compares the JM(1) and the SJM(1) results. These two
curves coincide at $\bar r_s\approx 4.18$ where the JM is mechanically stable.
The difference $W_{\rm SJM}-W_{\rm JM}$ is positive for $\bar r_s<4.18$ and
changes sign for $\bar r_s>4.18$. This behavior can easily be predicted if one
resorts to the corresponding expressions for the work functions: For
$\bar r_s\approx 4.18$, we have $\langle\delta v\rangle_{\rm WS}\approx 0$ and
therefore, the density profiles are the same which results in
$\Delta\phi[n_{\rm SJM}]=\Delta\phi[n_{\rm JM}]$ and in turn, by Eq.
(\ref{eq14}) leads to the same workfunctions. However, since the term
$\langle\delta v\rangle_{\rm WS}$ in Eq. (\ref{eq14}) changes sign at $\bar
r_s=4.18$, it gives rise to the change in the sign of $W_{\rm SJM}-W_{\rm JM}$.

In Fig. \ref{fig1}(d), the values of the quantum size correction $c$ for the
JM(1) and the
SJM(1) are compared. These two plots have two intersections. The one
at $\bar r_s\approx 4.18$
is obvious if we consider the dependence\cite{engper91,seidl98} of the quantity
$c$ on the density profile $n(x)$

\begin{equation}\label{eq17}
c=-h(-\infty)=4\pi\int_{-\infty}^{+\infty}dx\{x^2[n(x)-\bar
n\theta(-x)]+xf(x)\},
\end{equation}

\begin{eqnarray}
\int_{-\infty}^{+\infty}dxf(x)&=&-2\int_{-\infty}^{+\infty}dxx
[n(x)-\bar n\theta(-x)]  \\ \label{eq18}
     &=&-\frac{1}{2\pi}\Delta\phi  \label{eq19}.
\end{eqnarray}
The intersection at $\bar r_s\approx 2.5$ is because of the improper behavior
of the properties at high densities in the JM ( See for example Fig. 6(a) of
Ref. \ref{pay98} for the surface energy which becomes negative for $\bar
r_s<2.5$).

In Fig. \ref{fig2}(a), we have shown the position of the centroid of excess
charge, $a$, for the singly-ionized clusters as a function of $\bar r_s$ both
for
the JM(1) and the SJM(1). As is seen, both plots predict negative values which
means that the centroid for a positively charged cluster lies inside the
jellium [ the jellium surface is taken as the origin, see Eq. (\ref{eq6})].
This is consistent with our understandings that the excess charge in a metal
resides on the surface. On the other hand, the JM(1) and the SJM(1) have
opposite behaviors: the former is decreasing ( increasing in absolute value)
and the latter is increasing ( decreasing in absolute value) and intersect each
other at $\bar r_s\approx4.18$. The explanation of these opposite behaviors is
straightforward: In simple JM, if we take the thickness of the jellium shell at
the edge as $\lambda$, then the volume of this shell contains a unit
charge (for singly-ionized clusters)

\begin{equation}\label{eq20}
\lambda (4\pi R^2) \bar n=1,
\end{equation}
which gives

\begin{equation}\label{eq21}
a\approx -\frac{1}{2}\lambda=-\frac{1}{6R^2}\bar r_s^3.
\end{equation}
By Eq. (\ref{eq21}), for a fixed $R$, in JM as $r_s$ increases, the position of
the centroid moves inside from the edge towards the center of the jellium
sphere.
However, in the SJM, there exist two competing effects: On the one hand, as in
the simple JM, because the jellium background is the same, increasing $\bar
r_s$ has the tendency to increase $\lambda$ but, on the other hand, the SJM
Kohn-Sham effective potential contains an extra term
$\langle\delta v\rangle_{\rm WS}$ which is an increasing function of $\bar r_s$
( See Table IV of Ref. \ref{pertran} ). This term is negative for
$\bar r_s<4.18$ and hence deepens the effective potential ( compared to the JM
)
which reduces the spill-out of the electrons, so that $a_{\rm SJM}<a_{\rm JM}$.
On the other hand, for $\bar r_s>4.18$ the extra term is positive which
shallowen the
effective potential ( compared to the JM ) which in turn increases the range of
the Kohn-sham orbitals
from which the electron density is calculated. The result is that in the
SJM, the overlap of the negative and positive charges increases (the shell
thickness decreases) with increasing $\bar r_s$, so that $a_{\rm SJM}>a_{\rm
JM}$.

The quantity $a$ calculated above is, of course, the centroid position for
singly-ionized cluster. To calculate the centroid position for doubly-ionized
clusters, we fitted the Kohn-Sham energies of doubly-ionized clusters
to the Eq. (20) of Ref. \ref{seidl98} which is obtained
from Taylor expansion of the energy of a $z$-ply charged cluster about $z=0$.
The resulted values for $a$ was +0.34, -0.18, -0.29 for $\bar r_s$ values 2, 4,
6, respectively. Obviousely, the result for $\bar r_s=2$ is incorrect because
of its wrong sign. The predicted value for $\bar r_s=6$ is quite acceptable and
consistent with the above arguments. One reason for the incorrect results may
be in the use of Taylor expansion about $z=0$ for points $z>>0$, as in
Eq. (20) of Ref. \ref{seidl98}. A better approximation is obtained
by using the $z$-th ionization energy

\begin{eqnarray}
I_z(R)&=&E_R(N^*-z)-E_R(N^*-z+1) \\  \label{eeqq1}
&=&-\varepsilon_R^{\rm HO}(N^*-z+1)+\frac{1}{2(R+a_z)}, \label{eeqq2}
\end{eqnarray}
where we have used the Taylor expansion of the ground-state energy $E_R(N^*-z)$
at $N^*-z+1$. The quantity $\varepsilon_R^{\rm HO}(N^*-z+1)$ is the highest
occupied Kohn-Sham orbital energy\cite{janak} for the $(z-1)$-ply ionized
cluster with
radius $R=\bar r_s (N^*)^{1/3}$.

It is instructive to compare the behaviors of the position of the centroid of
excess charge with the position of the image plane of a planar surface. In Fig.
\ref{fig2}(b) we have compared these quantities in the JM and the SJM. As is
seen, both of them are positive and intersect at $\bar r_s\approx 4.$ In the JM
plot we have used the Eq. (8) of Ref. \ref{per88} and for the SJM we have
used the data from Fig. 1 of Ref. \ref{kiej93}. Here, although the argument
used for the finite cluster to calculate $\lambda$ is useless but, the effect
of
$\langle\delta v\rangle_{\rm}$ is the same as in the finite cluster case which
leads to similar behaviors.

Now, we show the results of fitting our self-consistent Kohn-Sham energies
to Eq. (\ref{eq15}) but neglecting
the quantum size correction, as in Eq. (16) of Ref. \ref{per88}.
Putting $c=0$ in Eq. (\ref{eq15}) we call the resulting equation as scheme 2.

In Fig. \ref{fig3}(a) we have compared the work functions obtained in schemes 1
and 2. As is seen, the scheme 2 predicts lower values than the scheme 1
otherwise the shapes are the same.

Figure \ref{fig3}(b) shows the plots of $a$ for JM and SJM in scheme 2. As we
see,
both of them are positive and have increasing behaviors. This means that the
position of the centroid of the excess positive charge (for the singly-ionized
clusters) lies outside the jellium edge which is incorrect because there exists
no positive charge. Therefore, inclusion of the quantum size effect is vital
for predicting correct values of the physical quantities.

Finally, we introduce the scheme 3 which is obtained by putting $a=0$ in Eq.
(\ref{eq15}) as Eqs. (1) of Refs. \ref{seidl98} and \ref{seidl97}.
In Fig. \ref{fig4}(a) we have compared the SJM work functions in the schemes 1
and 3 with the bulk values\cite{fiol92}. It is seen that scheme 3 improves the
workfunction and reduces the distance from the bulk values at higher densities.

The SJM fitted values of the size correction for schemes 1 and 3 are compared
in Fig. \ref{fig4}(b). It is seen that in scheme 3, the values are nearly
constant for the metallic densities $c\approx -3.7 eV$ in good agreement with
the result of Ref. \ref{seidl98}.

\section{summary and conclusions}
We have calculated the self-consistent Kohn-Sham energies for neutral and
singly-ionized clusters with different sizes and repeated the calculations for
different $\bar r_s$ values in the metallic range. The ionization energies for
each $\bar r_s$ has been separately fitted to Eq. (\ref{eq15}), we called
scheme 1. The results obtained for the position of the centroid of excess
charge is consistent with the understanding that excess charges reside at
the outer surface of a metal particle. However, for small
doubly-ionized clusters, using the Taylor expansion about $z=0$ gives incorrect
$a$ values for clusters of high electron-density metals. A better result would
be obtained by using a Taylor expansion of the energy of $z$-ply ionized
cluster about the point $(z-1)$.

Next, in scheme 2, we have fitted the same ionization energies
to Eq. (\ref{eq15})   taking $c=0$. The results show incorrect values for the
centroid position.

Finally,
in scheme 3, we put $a=0$ in Eq. (\ref{eq15}) and fit the ionization energies,
which results in better values for the work function at higher densities.

To conclude,
if one wishes to calculate the position of the centroid of excess charge of
$z$-ply charged cluster, one should use the Taylor expansion about point
$(z-1)$ as in Eq. (\ref{eeqq2}).
On the other hand, if one wishes choose one of the versions of Eq. (\ref{eq15})
to calculate the ionization
energies ( or electron affinities) of simple metal clusters, the best
choice would be the Eq. (\ref{eq15})
with the $\Delta SCF$ values for $W$
and $c$, and for sufficiently large clusters ($R>>a$) the scheme 3 with
the $\Delta SCF$ values for $W$ and $c$ gives the same results as scheme 2.

{\begin{center} \bf ACKNOWLEDGEMENT  \end{center}}

The author would like to thank Bahram Payami for providing computer facilities.

\newpage

\begin{figure}
\caption{Work functions and quantum size corrections in electron volts from
scheme 1. In (a)- the fitted values of the work function has been compared with
those of the bulk calculation for the SJM. (b)- the same as (a) but for the
JM. (c)- compares the fitted JM and SJM work functions. (d)- compares the
fitted quantum size corrections in the JM and the SJM. }
\label{fig1}
\end{figure}

\begin{figure}
\caption{(a)- The scheme 1 positions of the centroid of excess charge, in
atomic
units, as functions of $\bar r_s$ for the JM and the SJM. (b)- The positions of
the
image planes for the planar surfaces, in atomic units, from bulk calculations
of the JM and the SJM.}
\label{fig2}
\end{figure}

\begin{figure}
\caption{(a)- Work functions of schemes 1 and 2 in electron volts are compared.
(b)- Position of the centroid of excess charge, in atomic units, in scheme 2.
Both the SJM and the JM show increasing behaviors.}
\label{fig3}
\end{figure}

\begin{figure}
\caption{(a)- Work functions in electron volts for schemes 1 and 3 are
compared with the bulk. (b)- quantum size corrections in electrov volts for the
schemes 1 and 3. The scheme 3 predicts more or less a constant value.}
\label{fig4}
\end{figure}

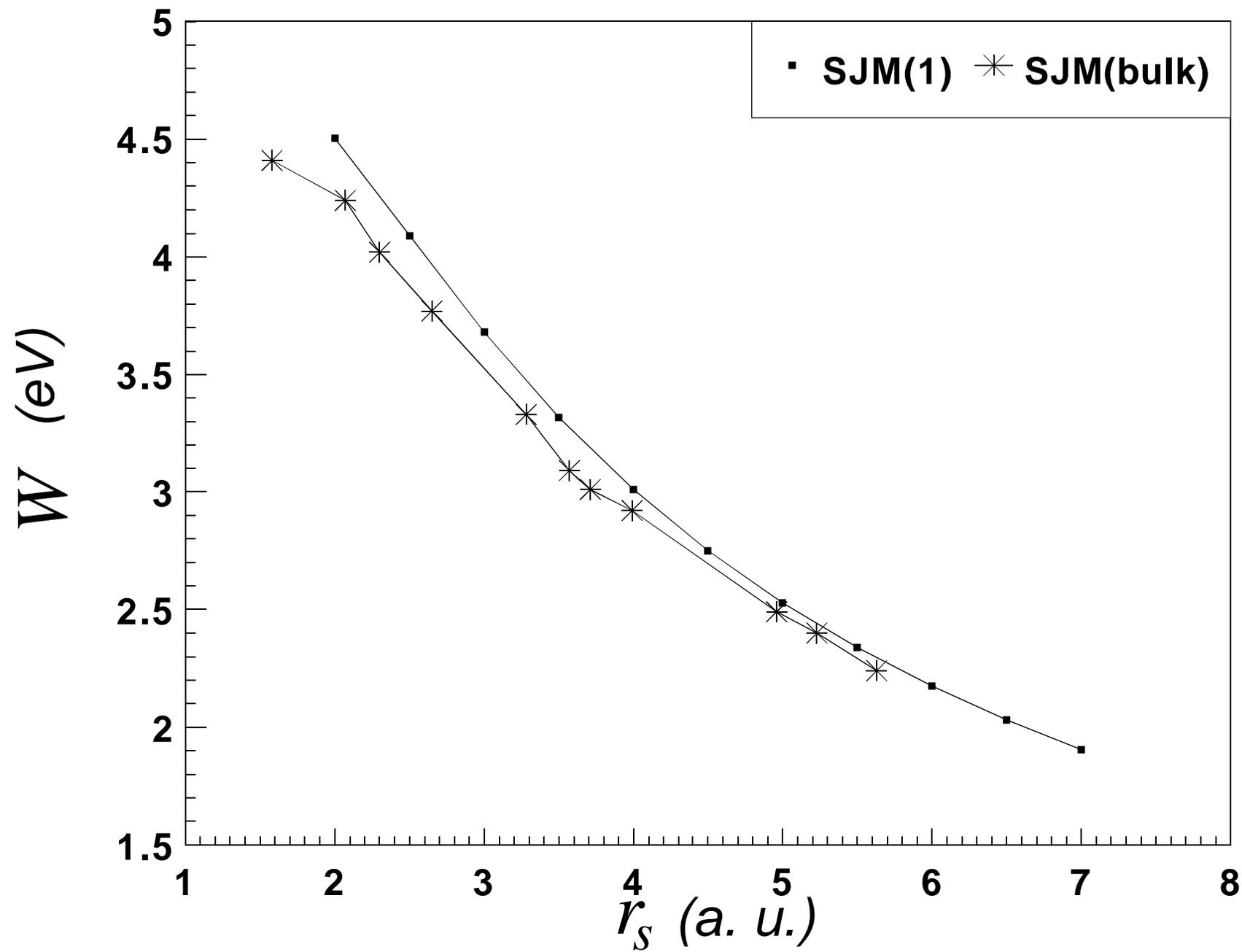
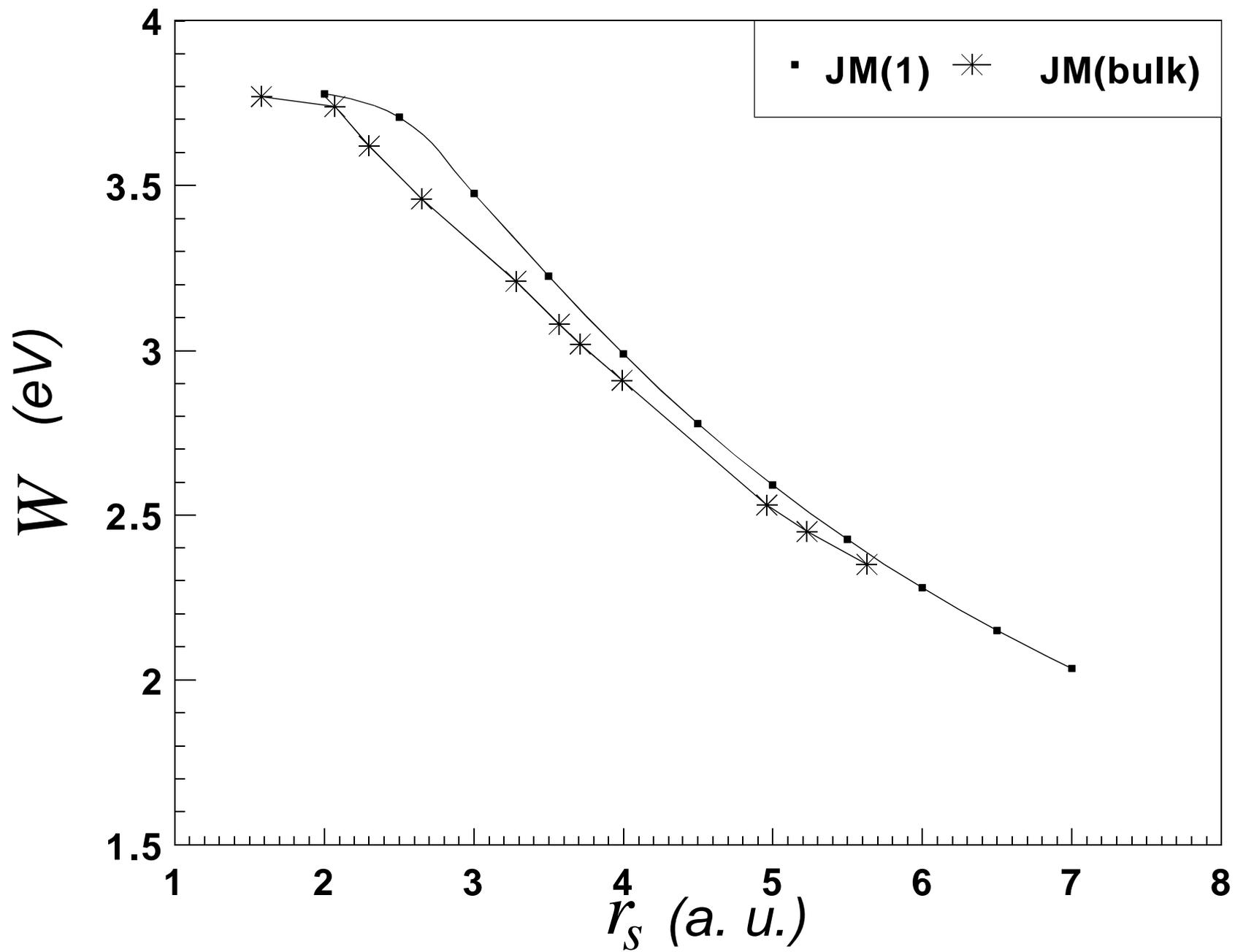
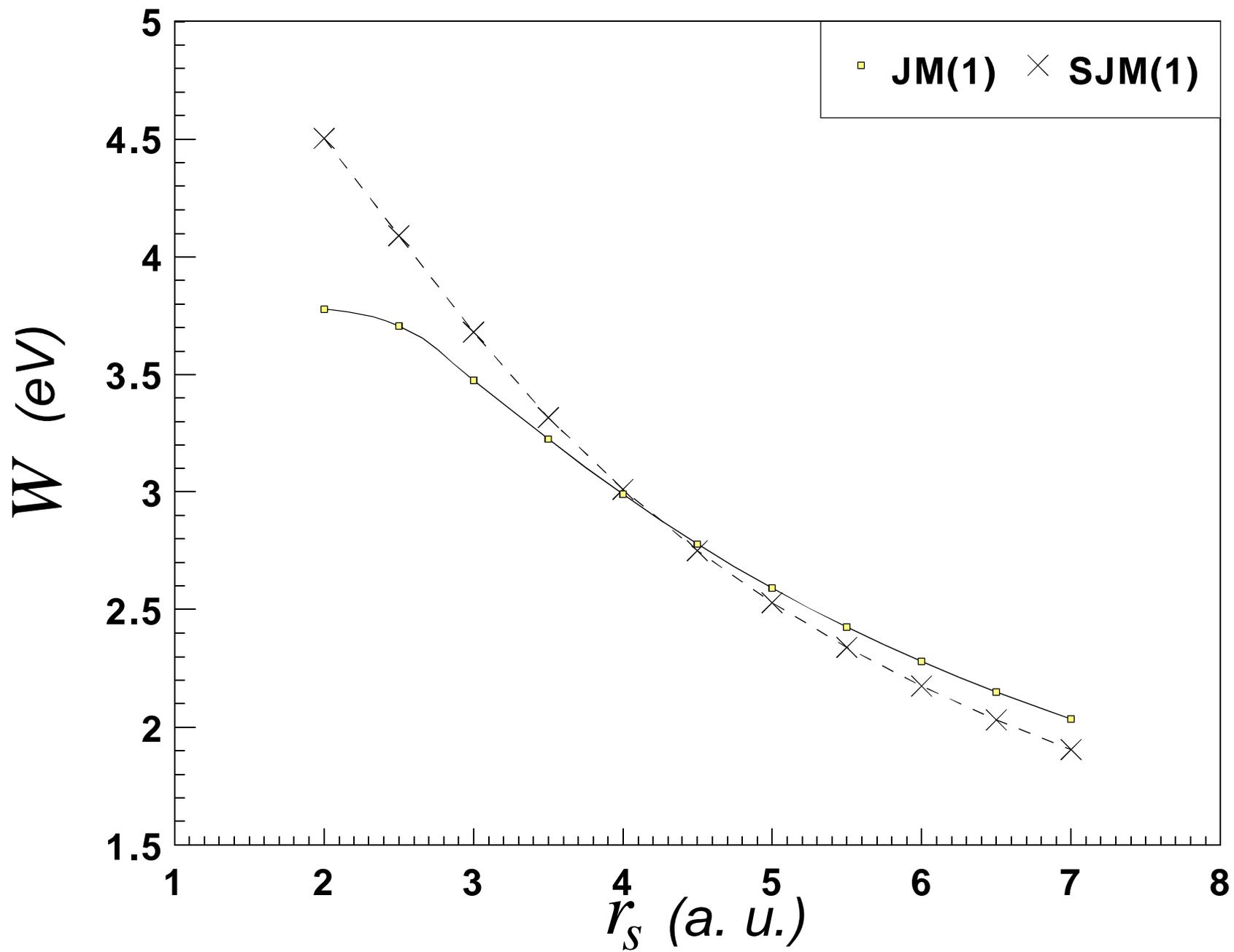
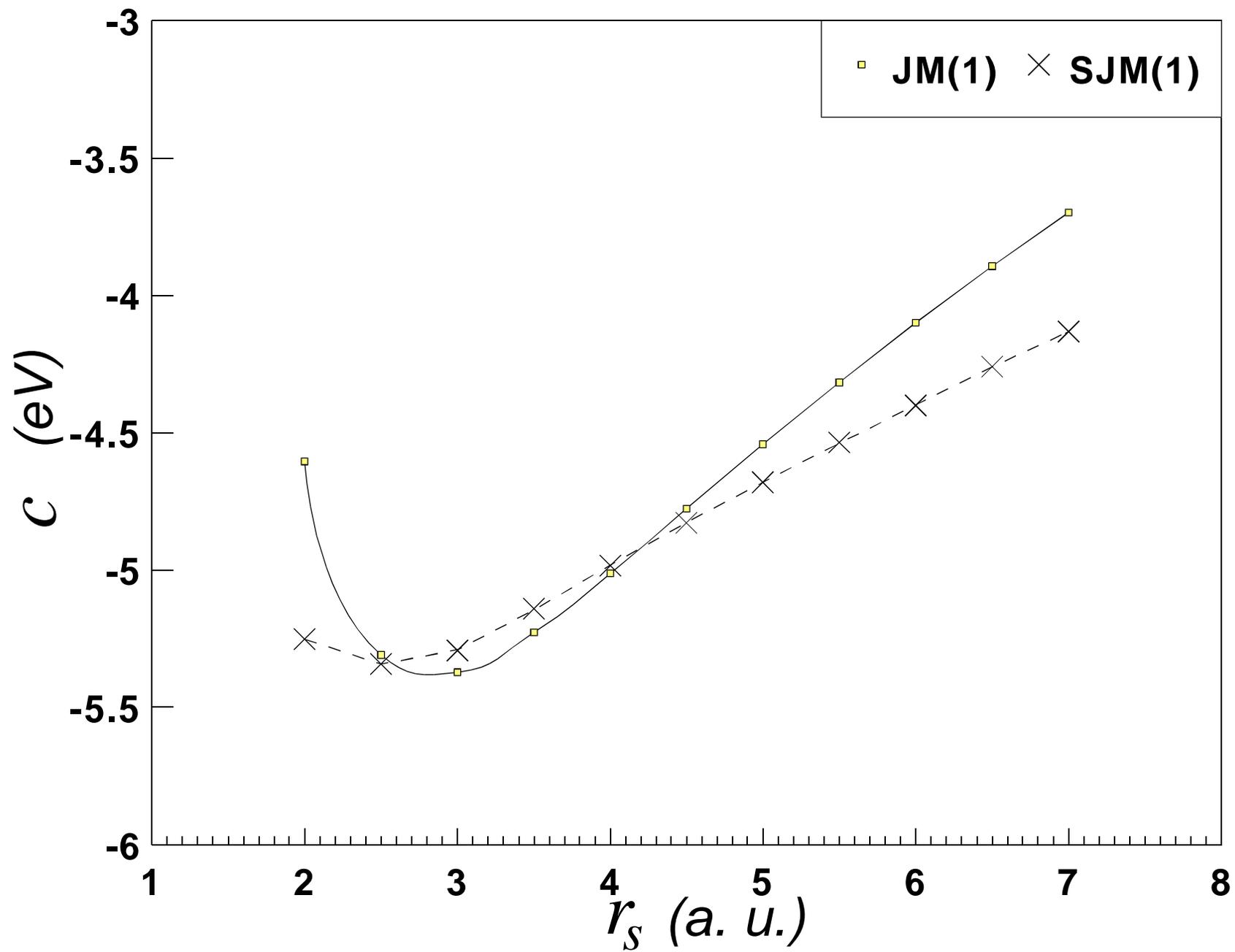
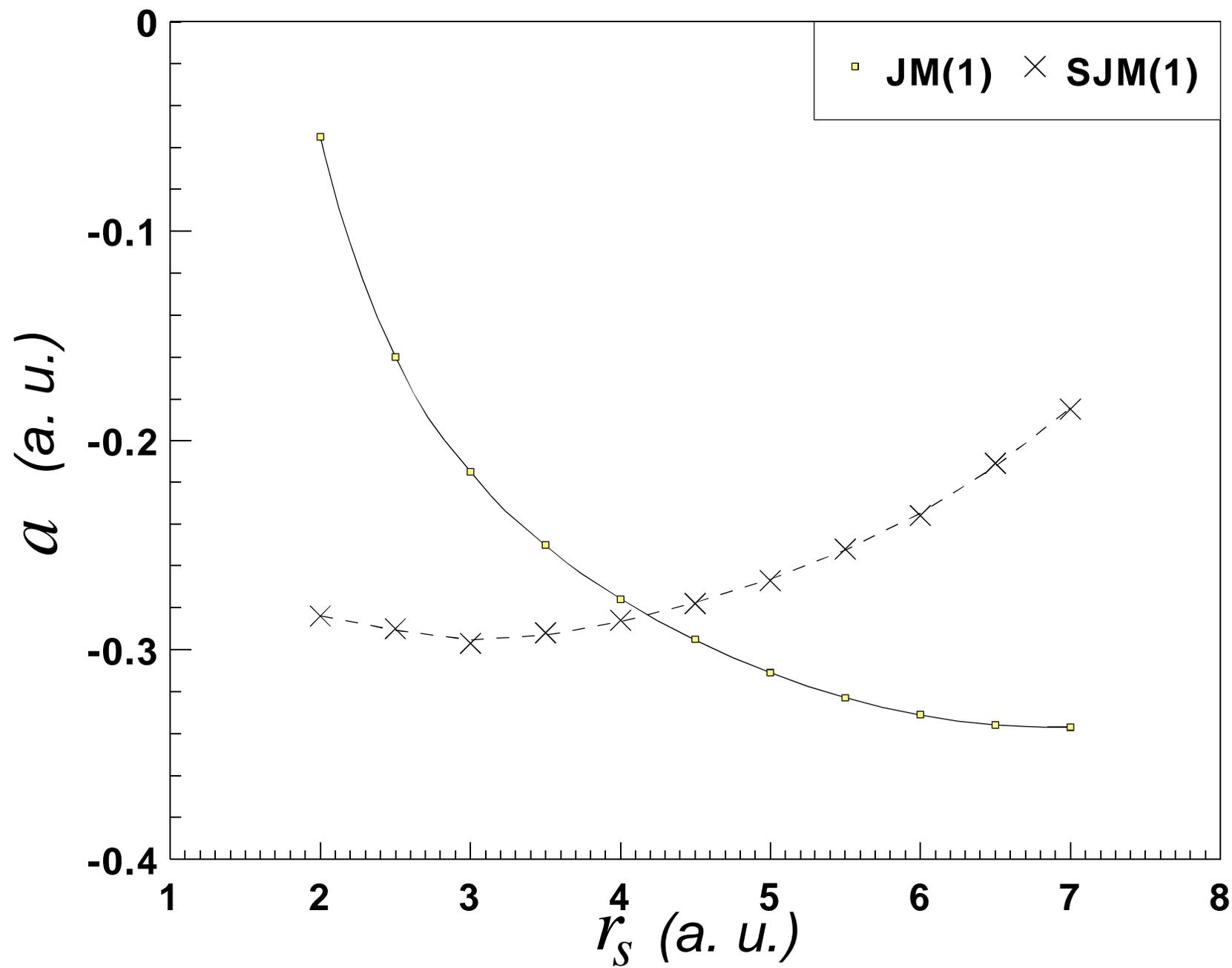
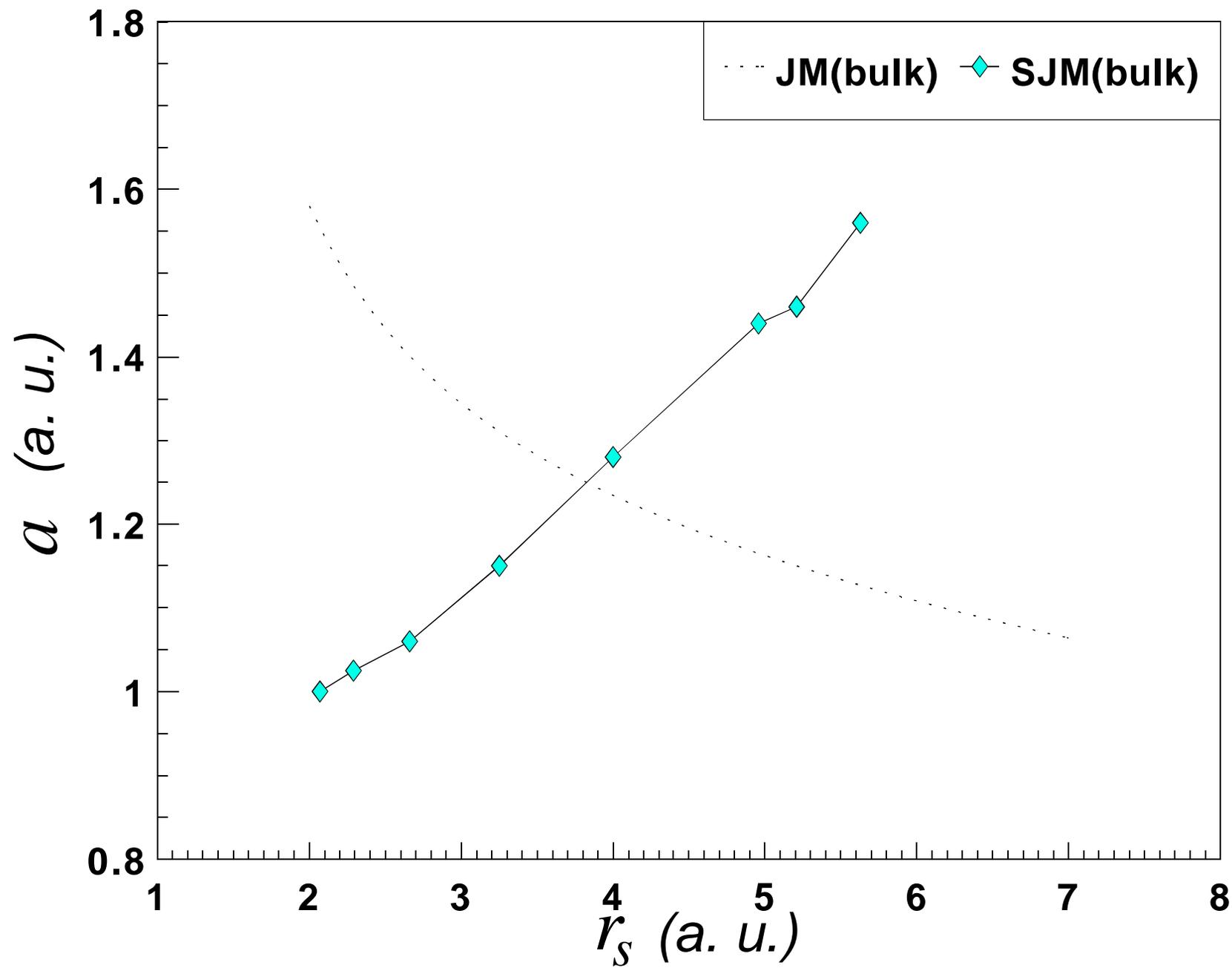
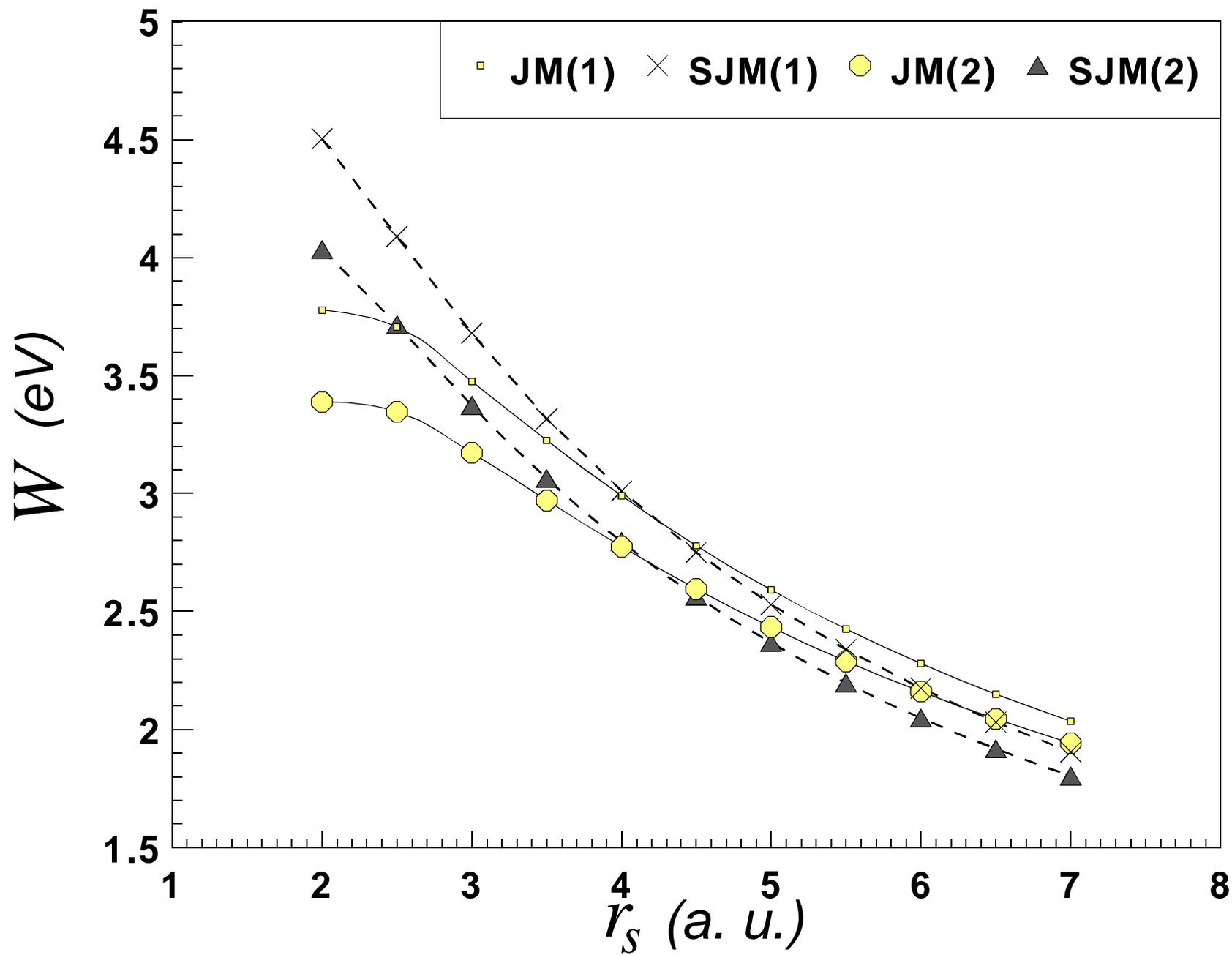
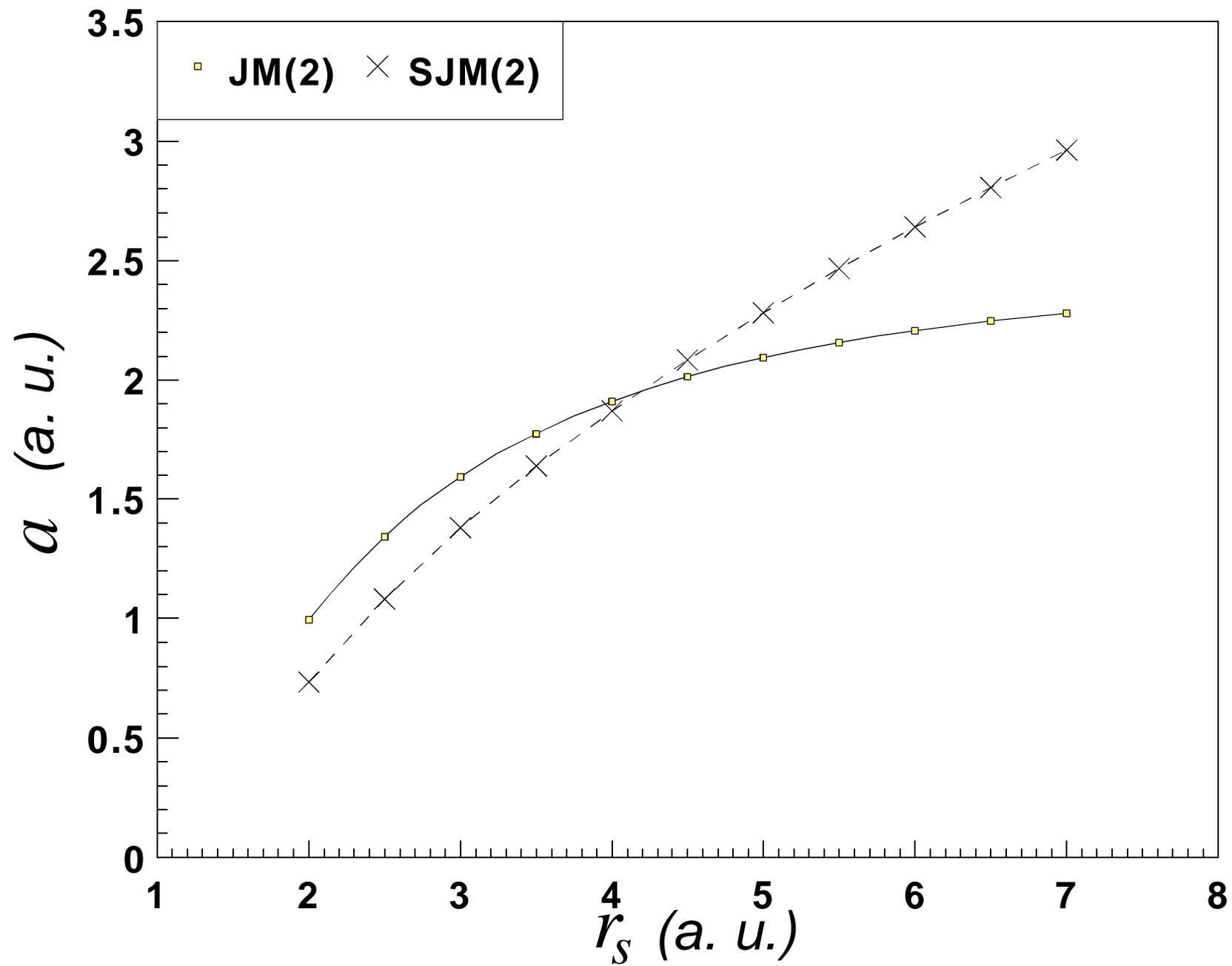
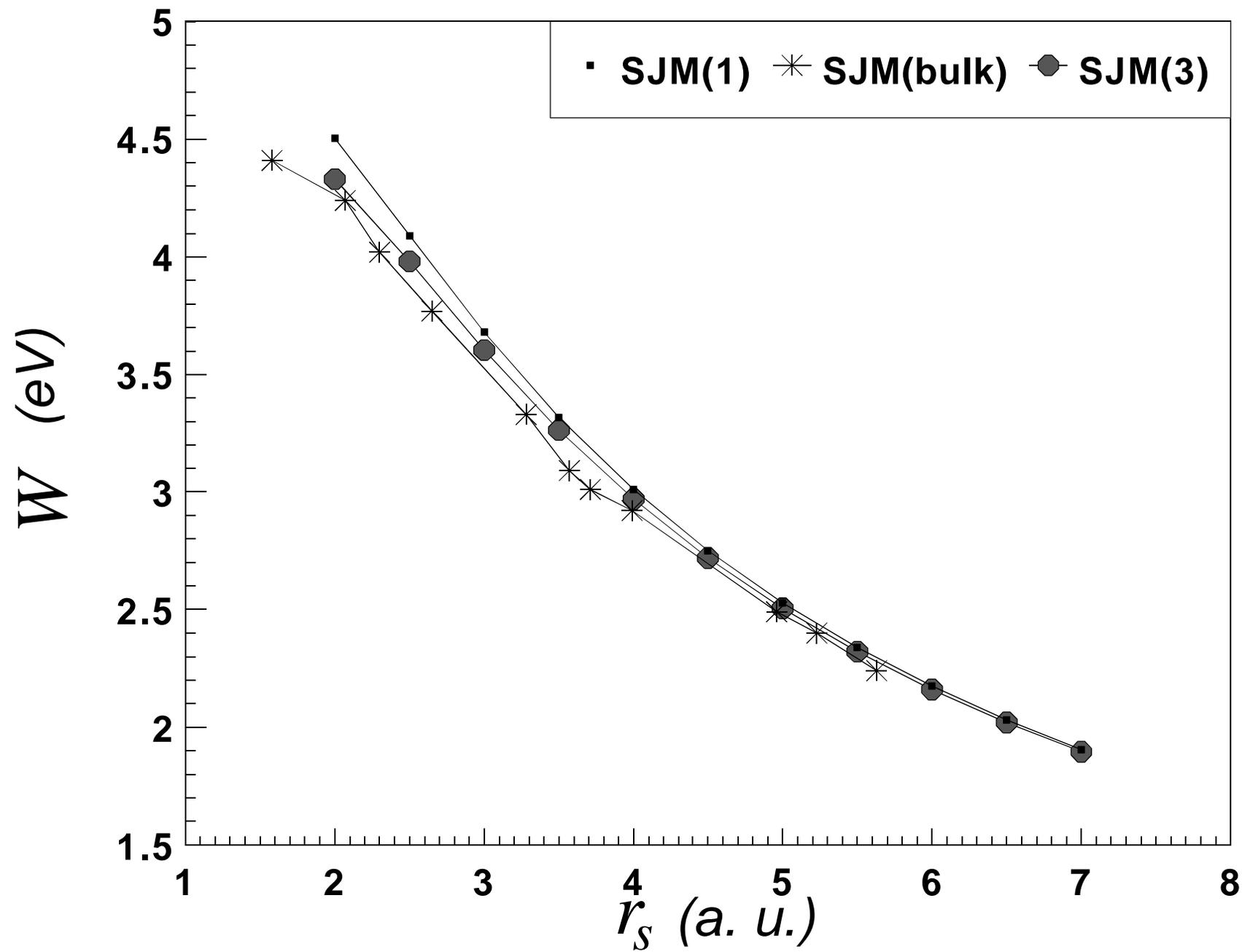
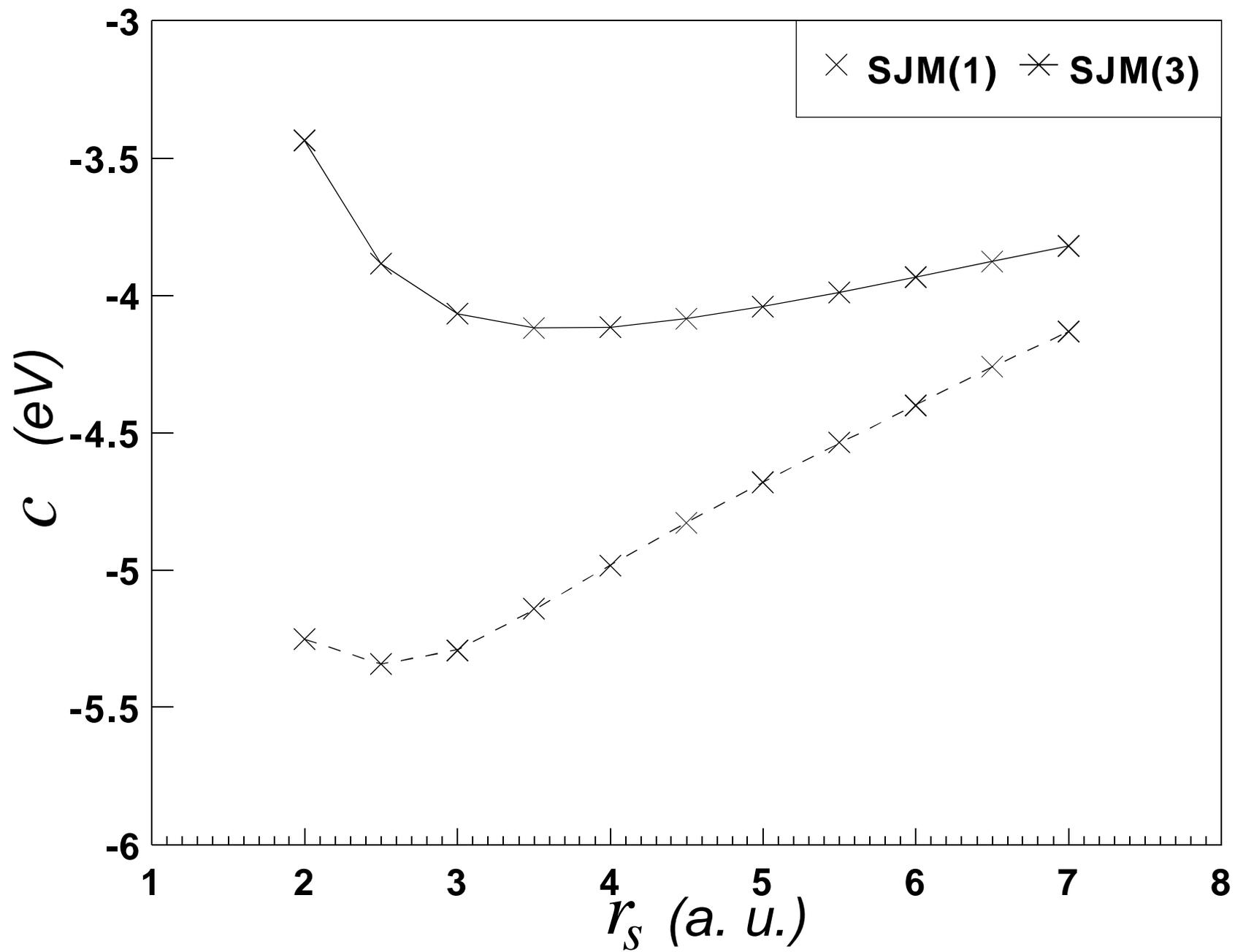
\end{document}